\documentstyle[aps,multicol,epsf,psfig]{revtex}

\newcommand{\beq}{\begin{equation}}
\newcommand{\eeq}{\end{equation}}
\newcommand{\beqa}{\begin{eqnarray}}
\newcommand{\eeqa}{\end{eqnarray}}

\def\opone{\leavevmode\hbox{\small1\kern-3.8pt\normalsize1}}

\begin{document}
\title{Classical Teleportation of a Quantum Bit}
\author{N. J. Cerf$^1$, N. Gisin$^2$, and S. Massar$^3$}
\address{$^1$ Ecole Polytechnique, CP 165, Free University of Brussels,
1050 Brussels, Belgium\\
$^2$ Group of Applied Physics, University of Geneva, 1211 Geneva, Switzerland\\
$^3$ Service de Physique Th\'eorique, CP 225, Free University of Brussels,
1050 Brussels, Belgium}

\date{28 June 1999}
\draft
\maketitle

\begin{abstract}
Classical teleportation is defined as a scenario where the sender
is given the classical description of an arbitrary quantum state 
while the receiver simulates any measurement on it. 
This scenario is shown to be achievable by transmitting only 
a few classical bits if the sender and receiver initially share
local hidden variables. Specifically, a communication of 2.19~bits
is sufficient on average for the classical teleportation of a qubit, 
when restricted to von Neumann measurements. The generalization 
to positive-operator-valued measurements is also discussed.
\end{abstract}

\pacs{PACS numbers: 03.65.Bz, 03.67.-a, 89.70.+c
\hfill preprint ULB-TH/99-11}

\vspace{-0.4cm}

\begin{multicols}{2}

The past few years have seen the emergence of quantum information
theory. This generalization of Shannon's classical theory of
information describes how quantum systems can be manipulated
from an information-theoretic perspective. In particular, whereas the basic
quantity of Shannon's theory is the amount of information
that can be stored in a two-state register (a bit), 
the basic quantity of quantum information theory is the amount of 
information that is contained in a quantum system whose
Hilbert space is two-dimensional (called a qubit). 
Also, quantum information theory deals with completely novel
properties that have no classical counterpart
such as quantum entanglement (non-local correlations),
and their role in communication problems. 
\par

The exploration of the relation between classical and quantum
informational resources is a central objective of this new discipline. 
Perhaps one of the most celebrated example of such a relation is
{\em quantum teleportation}\cite{BBCJPW93}, 
in which the physical transmission of a qubit is replaced by 
the transmission of 2 classical bits
supplemented with preexisting entanglement. The analysis of
this process raises a fundamental question: What is the
exact information content carried by a qubit?
On the one hand, specifying a pure state in a 2-dimensional
Hilbert space, i.e., $|\psi\rangle = \cos{\theta\over 2}|0\rangle
+ e^{i\phi} \sin{\theta\over 2} |1\rangle$, requires two real numbers 
(the angles $\theta$ and $\phi$ in the Bloch sphere), 
so that communicating a {\em known} quantum state apparently 
requires an {\em infinite} number of classical bits. 
On the other hand, it happens that 
teleportation can be achieved by transmitting only 2 classical bits,
provided that they are supplemented with a previously existing 
entangled pair of qubits. Yet, as such, an entangled pair 
cannot do more than generate correlated random numbers, and surely
not be used to communicate---this would violate causality.
This state of affairs led Vaidman to question 
{\it ``whether the essence of a quantum state 
is only 2~bits?''}\cite{Vaidman98}
\par

In this Letter, we shall show that the situation is even more surprising 
than envisaged by Vaidman. We shall exhibit a scenario, denoted
as {\em classical teleportation}, in which a quantum entangled
pair is not even necessary, and the teleportation of a {\em known} qubit
can be simulated entirely classically. The sender, called Alice, 
is given the classical description of a quantum state, i.~e., 
a vector $\vec a$ on the Bloch sphere. The receiver, Bob, 
chooses a particular measurement, that is, a vector $\vec b$ on
the Bloch sphere in the case of a von Neumann measurement. 
Importantly,
Bob's choice is unknown to Alice, and Bob is ignorant of Alice's 
quantum state. Bob's task is to give an outcome
that is consistent with him having performed a measurement
on Alice's state. Here, we will show that if 
Alice and Bob share initially some local ``hidden'' variables 
(LHV), that is, they possess an identical (possibly infinite) list 
of random numbers, then carrying out classical teleportation requires
only a finite---and very limited---amount of communication. 
In other words, all possible (von Neumann) measurements performed by Bob
can be simulated by Alice transmitting a few classical bits to Bob,
{\em as if} the qubit had actually been teleported.
We find that, if Bob restricts himself
to von Neumann measurements, an average of 2.19~bits
of classical communication is sufficient to complete this task, 
which is only 0.19~bits more than the amount needed for 
quantum teleportation. More generally,
if Bob's measurement is based on a Positive-Operator-Valued Measure (POVM),
the best algorithm we have found is less efficient
and uses a total of 6.38 bits of two-way communication.
\par

Of course, this restricted scenario cannot simulate all aspects of the
teleportation of a qubit; in particular, entanglement with respect
to another system cannot be transmitted. Instead, we show that
Bob can simulate the statistics of any measurement on a quantum
state known by Alice but unknown to him,
even though Alice and Bob do not share prior entanglement. 
Moreover, it should
be noted that achieving classical teleportation using one-way
classical communication {\em but} with no shared LHVs necessarily
requires the transmission of an infinite number of bits (Alice should
then transmit to Bob the description of her state, that is, the
angles $\theta$ and $\phi$). 
\par

The protocol we shall present below is inspired by, and closely related 
to schemes for classically simulating Bell correlations 
recently discovered by Brassard {\it et al.} \cite{BCT99}, 
and independently by Steiner\cite{Steiner99}. It is well known that
any local hidden-variable model of quantum mechanics {\em cannot} reproduce
quantum correlations, as reflected by the violation of Bell inequalities.
Surprisingly, these authors have demonstrated that
supplementing a local hidden-variable model with a very limited amount of 
classical communication suffices to perfectly simulate quantum
correlations. Our protocol for the classical teleportation of 
a qubit is based on the
generalization of Steiner's protocol to the Bloch sphere (Steiner's protocol
was restricted to {\em real} quantum mechanics, i.e., $\phi=0$)
presented in Ref.~\cite{gisingisin}, 
where it was used to analyze the detection loophole in Bell experiments.
It differs from that scheme by the use of {\em triplets} of orthogonal vectors 
so as to span the Bloch sphere more uniformally.
It also decreases significantly the amount of communication needed 
by better exploiting the shared variables between Alice and Bob.
\par

Before presenting the details of our classical scenario, 
let us highlight its central features.
Initially, Alice and Bob share an infinite list 
of triplets of normalized vectors distributed randomly on the 
Bloch sphere, $\{\vec\lambda_k,\vec\mu_k,\vec\nu_k\}$, with $k=1,2,\ldots$.
For each $k$, the three vectors form an
orthogonal 
frame oriented randomly and chosen independently of the other triplets.
In addition, Alice and Bob also possess a common infinite list 
of independent random numbers $u_k$ (with $k=1,2,\cdots$)
uniformly distributed in the interval $[0,\sqrt{3}]$. Together the
infinite set of ($\vec\lambda_k,\vec\mu_k,\vec\nu_k,u_k$) constitute
the LHVs~\cite{fn1}.
Once Alice and Bob are separated, Alice is given 
an arbitrary normalized vector $\vec a$ on the Bloch sphere 
specifying the quantum state $|\psi(\vec a)\rangle$
to be communicated to Bob, 
while Bob chooses an arbitrary von Neumann measurement (described
by vector $\vec b$). Each party is oblivious of the
choice the other party has made.
Thus, the quantum state chosen by Alice (and unknown to Bob)
is characterized by the projector 
$P_{\vec a}=(\openone+\vec a \cdot \vec \sigma)/2$, where
$\vec\sigma$ denotes the Pauli matrices. 
The measurement chosen by Bob (unknown to Alice) corresponds 
to the observable $\sigma_{\vec b}=\vec b \cdot \vec\sigma$, resulting in
outcome $\pm$ with probability
\beq  \label{qmstat}
p(\pm|\vec a)={\rm Tr} \left( P_{\pm\vec b} \, P_{\vec a} \right) 
= {1\pm \vec a\cdot \vec b \over 2}
\eeq
where $P_{\pm \vec b} =(\openone\pm \vec b \cdot \vec \sigma)/2$.
The goal is for Bob to generate measurement statistics 
{\em identical} to that predicted by quantum mechanics, Eq.~(\ref{qmstat}),
but without Alice actually transmitting her qubit to Bob.
Instead, the variables ($\vec\lambda_k,\vec\mu_k,\vec\nu_k,u_k$)
are supplemented with classical communication.
In addition, we will assume that the scenario
is repeated many times, that is, Alice prepares $N$ qubits
in states $|\psi(\vec a_1)\rangle, |\psi(\vec a_2)\rangle, \cdots,
|\psi(\vec a_N)\rangle$, and sends them to Bob, whereupon Bob
carries out separate measurements $\sigma_{\vec b_1},\sigma_{\vec b_2},
\cdots,\sigma_{\vec b_N}$ on each qubit. The transmission of $N$ qubits
is simulated 
in parallel, allowing the protocol to make use
of 
block coding, hence minimizing the amount of classical
communication needed.
\par

Our protocol works as follows. 
First, Alice divides the interval $[0,\sqrt{3}]$ of each $u_k$ 
into 4 segments. These segments depend on the orientation of the
corresponding variables $(\vec\lambda_k,\vec\mu_k,\vec\nu_k)$ 
with respect to $\vec a$ as
\beqa
&\mbox{Zone A$_\lambda$} :& 0 \le u_k < |\vec a \cdot \vec\lambda_k|
\nonumber\\
&\mbox{Zone A$_\mu$} :& 0 \le u_k - |\vec a \cdot \vec\lambda_k| <
                   |\vec a \cdot \vec\mu_k|
\nonumber\\
&\mbox{Zone A$_\nu$} :& 0 \le u_k - |\vec a \cdot \vec\lambda_k|
   -|\vec a \cdot \vec\mu_k|  < |\vec a \cdot \vec\nu_k|
\nonumber\\
&\mbox{Zone R\phantom{$\nu$}} :& |\vec a \cdot \vec\lambda_k|
   +|\vec a \cdot \vec\mu_k| + |\vec a \cdot \vec\nu_k| \le u_k <\sqrt{3}
\label{2}
\eeqa
Alice then performs the following operations:
\begin{description}
\item[A1]
She sets the index $k=1$.
\item[A2]
She checks whether $u_k$ belongs to zone R. If it
does, she {\em rejects} the triplet $(\vec\lambda_k,\vec\mu_k,\vec\nu_k)$, 
that is, she increments $k$ by one 
and goes back to step {\bf A2}.
\item[A3]
Once the $k$th triplet has been {\em accepted},
Alice communicates to Bob at which iteration $u_k$ first 
belonged to one of the zones A$_\lambda$, A$_\mu$, or A$_\nu$, and
to which of these zones it belongs. 
\item[A4] 
She also sends an extra bit which is 
the sign of $\vec a \cdot \vec\lambda_k$
if $u_k \in {\rm A}_\lambda$, or similarly for A$_\mu$ and A$_\nu$.
\end{description}
Upon receiving this information from Alice, Bob knows which
vector $\vec\lambda_k$, $\vec\mu_k$, or $\vec\nu_k$,
was accepted by Alice. Denote this vector by $\vec \lambda$. He then
carries out the following operations:
\begin{description}
\item[B1]
He flips $\vec \lambda \to -\vec\lambda$
if the sign of $\vec a \cdot \vec\lambda$ is negative.
\item[B2]
The outcome $\pm$ of Bob's measurement of $\vec \sigma_{\vec b}$ is
then given by the sign of $\vec b \cdot \vec \lambda$.
\end{description}
\par

Let us check that, with this protocol, Bob correctly reproduces the
quantum probabilities. Since the random variable $u_k$ is uniformly
distributed, the probability of accepting $\vec\lambda_k$ conditionally
on $\vec a$, i.e., 
$p(\vec\lambda_k|\vec a)=|\vec a \cdot \vec \lambda_k|/(4\pi\sqrt{3})$.
Using the isotropy of the $\vec\lambda_k$-distribution,
the average probability of acceptance (in zone $A_\lambda$) is given by
\beq
p_{A_\lambda}= \int d\vec\lambda \;
p(\vec\lambda|\vec a)={1\over 2\sqrt{3}}
\eeq
at each iteration $k$. Note that it does not depend on $\vec a$.
Similarly, we have $p(\vec\mu_k|\vec a)=|\vec a \cdot \vec \mu_k|/
(4\pi\sqrt{3})$
and $p(\vec\nu_k|\vec a)=|\vec a \cdot \vec \nu_k|/(4\pi\sqrt{3})$,
so that $p_{A_\mu} = p_{A_\nu} = 1/( 2\sqrt{3})$.
Thus, the average probability of acceptance (in any zone) is
$p_A=p_{A_\lambda}+p_{A_\mu}+p_{A_\nu}=\sqrt{3}/2$, while
the corresponding average probability of rejection
is simply $p_R=1-p_A=(2-\sqrt{3})/2$. 
As far as the statistics of Bob's
outcome is concerned, it is sufficient to consider one of the
vectors $\vec\lambda_k$, $\vec\mu_k$, and $\vec\nu_k$, since
the correlation of $\vec \lambda$ with $\vec a$ expressed by
$p(\vec \lambda|\vec a) =|\vec a\cdot \vec \lambda|/(4\pi\sqrt{3})$ 
is independent of whether 
$\vec \lambda = \vec\lambda_k$, $\vec \mu_k$, or $\vec\nu_k$.
Once $\vec\lambda$ has been accepted, 
its {\it a posteriori} probability distribution is given by
\beq
P({\rm accepted~}\vec\lambda|\vec a)
= { p(\vec\lambda|\vec a) \over p_{A_\lambda} }
={|\vec a \cdot \vec \lambda| \over 2\pi}
\eeq
Consequently,
the probability that Bob chooses outcome $\pm$ for a given $\vec a$ is
\beqa
\lefteqn{ p(\pm|\vec a) = 
\int d\vec\lambda \; P({\rm accepted~}\vec\lambda|\vec a) } 
\hspace {0.5 cm} \nonumber\\
& & \times \big[ \Theta(\vec a \cdot \vec \lambda) 
             \, \Theta(\pm\vec b \cdot \vec \lambda)
      +  \Theta(-\vec a \cdot \vec \lambda) 
             \, \Theta(\mp\vec b \cdot \vec \lambda) \big]
\eeqa
where $\Theta(\cdot)$ is the Heaviside step function.
To verify that this integral coincides with the predictions 
of quantum mechanics, Eq.~(\ref{qmstat}),
it is easiest to calculate $p(+|\vec a) - p(-|\vec a)
-p(+|-\vec a) + p(-|-\vec a)$ and use the symmetry of the problem
(see \cite{gisingisin}).  
\par

We now compute the amount of classical information that must be communicated
by Alice to Bob in order to carry out this protocol. It will appear
that the use of three vectors $\vec\lambda_k$,
$\vec\mu_k$, and $\vec\nu_k$ is helpful to reduce this amount.
Ignoring the extra bit (the sign of $\vec a \cdot \vec \lambda$)
for the moment, the information needed by Bob to choose an outcome is 
the value of $k$ (the iteration at which a 
triplet has been accepted) and whether $\vec\lambda_k$,
$\vec\mu_k$, or $\vec\nu_k$ was accepted. Denote the latter
variable as $l=1,2,3$.
If block coding is used, the minimum number of bits 
that must be communicated per simulation of one qubit
is given asymptotically by the Shannon entropy of $k$ and $l$.
A crucial point is that the random number $u_k$ 
used by Alice to accept or reject the $k$th triplet
is also known to Bob, so that the latter can infer the {\it a priori}
probabilities of Alice choosing
$\vec\lambda_k$, $\vec\mu_k$, or $\vec\nu_k$.
Therefore, the {\em minimum} amount of information
that must be communicated on average
is the entropy of $k$ and $l$ {\em conditional} on the infinite set of
variables $u_k$'s, that is, the uncertainty on $(k,l)$ when
$u_1,u_2,\cdots$ are known.
\par

To calculate this quantity, note that, 
for a given set $u_1,u_2,\cdots$, the probability of accepting $A_l$
at the $k$th iteration is given by
\beq   \label{pkl}
p(k,l|u_1,u_2,\cdots)=p(R|u_1)\cdots p(R|u_{k-1}) p (A_l |u_k)
\eeq
where $p(A_l|u)$ is the {\it a priori} 
probability of accepting $A_l$ knowing $u$, 
and $p(R|u)$ is the {\it a priori} 
probability of rejection for a given $u$. 
For instance, as
$\vec\lambda$ is uniformly
distributed on the Bloch sphere, we have
\beq
p (A_\lambda |u)=\int {d\vec\lambda \over 4\pi} \,
\Theta\left( |\vec a \cdot \vec\lambda|-u \right)
\eeq
Similar expressions can be obtained from Eq.~(\ref{2}) for 
$p (A_\mu | u)$, $p (A_\nu | u)$, and $p (R|u)$. 
Since these four probabilities are not equal, 
block coding allows Alice to compress the information sent to Bob.
The entropy of $k$ and $l$ conditional on
$u_1,u_2,\cdots$ (averaged over these variables) is given by
\beqa
\lefteqn{ H\equiv H(k,l|u_1,u_2,\cdots) 
= - \int_0^{\sqrt{3}} {du_1\over \sqrt{3}} 
\int_0^{\sqrt{3}} {du_2\over \sqrt{3}} \cdots }
\nonumber   \\
&\times& \sum_{k=1}^\infty \sum_{l=1}^3
p(k,l|u_1,u_2,\cdots)\; \log_2 p(k,l|u_1,u_2,\cdots)
\eeqa
Using Eq.~(\ref{pkl}), we get
\beqa
H = \sum_{k=1}^\infty \sum_{l=1}^3
\left[ (k-1) p_R^{k-2}\, p_{A_l}\, q_R + p_R^{k-1}\, q_{A_l} \right]
\label{9}
\eeqa
where
\beqa
p_{A_l}&=&\int_0^{\sqrt{3}} {du \over \sqrt{3}}\; p_{A_l}(u)  \nonumber\\
q_{A_l}&=&-\int_0^{\sqrt{3}} {du \over \sqrt{3}}\; p_{A_l}(u)
\log_2 p_{A_l}(u)
\eeqa
and similarly for $p_R$ and $q_R$.
Carrying out the sum in Eq.~(\ref{9}), we obtain the entropy
$H = (q_A + q_R) / p_A$, with $p_A = \sum_{l=1}^3 p_{A_l}= \sqrt{3}/2$ and
$q_A= \sum_{l=1}^3 q_{A_l}$.
The expressions for $q_{A_l}$ and $q_R$ can be estimated numerically, giving 
$q_{A_\lambda}=0.207~$bits, $q_{A_\mu}=0.366~$bits, 
$q_{A_\nu}=0.341~$bits, and $q_{R}=0.117~$bits. 
This results in $H=1.19~$bits for the amount of information
needed for Bob to know which $\vec \lambda$ was accepted.
Consequently,
taking into account the extra bit needed to communicate
the sign of $\vec a \cdot \vec \lambda$,
the total amount of classical information that must be
transmitted is 1.19+1=2.19~bits, as announced.

This scheme can also be applied with minor modifications to the simulation of 
Bell correlations in a singlet state, as in
Refs.~\cite{BCT99,Steiner99,gisingisin}.
The relation is the following: suppose Alice and Bob 
have a classical protocol for simulating measurements on a singlet,
as  in
Refs.~\cite{BCT99,Steiner99,gisingisin}. If
Alice sends Bob an additional bit telling him which was the outcome
of her measurement, then this is equivalent
to Alice simulating the teleportation of 
a known qubit to Bob. Indeed, if Bob flips the outcome of his measurement
conditionally on the outcome of Alice's measurement of $\vec \sigma_{\vec a}$,
it is as if he had received the state $|\psi(\vec a)\rangle$
as a consequence of the exact anticorrelations exhibited by a singlet. 
The protocol we described above is equivalent to
Alice simulating the measurement of $\sigma_{\vec a}$ on a singlet,
supplemented by the transmission of one extra bit 
identifying which outcome she obtained. Hence, the simulation of the
singlet with our scheme necessitates only 1.19~bits on average.
For comparison, the protocol of Brassard {\it et al.} \cite{BCT99} uses 
exactly 8~bits, Steiner's \cite{Steiner99} uses 2.97~bits on average 
(when adapted to general von Neumann measurements), whereas the protocol 
of Ref.~\cite{gisingisin} uses 2~bits on average.
Also, Cleve \cite{Cleve99} proposed a protocol requiring slightly 
less than 2 bits on average.
\par

The most general measurement that Bob can carry out on a qubit is 
based on a POVM. We will now show that such a measurement can also be
simulated, at the cost of a larger amount of (two-way) communication.
The elements $B_j$ of Bob's POVM (with $\sum_j B_j = \openone$)
are taken to be proportional to
one-dimensional projectors (other POVM's can trivially be obtained
from such maximally refined POVM's). They can be expressed in
terms of vectors $\vec b_j$ in the Bloch sphere as 
$B_j = (|b_j| \openone + \vec b_j \cdot \vec \sigma )/2$, 
with the completeness conditions $\sum_j |b_j| =2$ and 
$\sum_j  \vec b_j  =0$.
The protocol starts by Alice carrying out her part as in steps {\bf A1}
to {\bf A4} above. We still assume that many qubits are teleported 
in parallel
so that block coding can be used.
Bob then performs the following operations:
\begin{description}
\item[B1'] Same as {\bf B1}. 
\item[B2'] 
He chooses randomly the $j$th outcome of his POVM
with probability $|b_j|/2$.
\item[B3'] 
He checks whether outcome $j$ is consistent with the 
accepted vector $\vec \lambda$ that was chosen by Alice. To this end, 
he computes the scalar product of $\vec \lambda$ with $\vec b_j$. 
If $\vec \lambda \cdot \vec b_j \ge 0$, 
he accepts outcome $j$ and sends a bit 0 back to Alice to let her know  
that he has chosen an outcome.
\item[B4'] 
If $\vec \lambda \cdot \vec b_j<0$, he then sends a bit 1 
to Alice to inform her that he was unable to choose an outcome. 
\end{description} 
If Alice receives a bit set to 0, she terminates. But if she receives a 
bit set to 1, she increments $k$ by 1 and returns to step {\bf A2} of her 
part of the protocol. Alice and Bob iterate this two-way protocol until Bob
can choose an outcome.
\par

Thus, the probability that Bob accepts $\lambda$ and gives
outcome $j$ is
\beqa
P(j|\vec a) &=&
{|b_j | \over 2} \int d\vec\lambda \; P({\rm accepted~}\vec\lambda|\vec a)
\nonumber \\
&&\times \big[ \Theta(\vec a \cdot \vec \lambda) 
                \, \Theta(\vec b_j \cdot \vec \lambda)
             + \Theta(-\vec a \cdot \vec \lambda) 
                \, \Theta(-\vec b_j \cdot \vec \lambda)  \big]
\nonumber\\
&=& {|b_j| + \vec a \cdot \vec b_j \over 4}
\label{13}
\eeqa
Hence, the probability that Bob terminates at step {\bf B3'} (averaged 
over $j$) is $\sum_j P(j|\vec a)= 1/2$.  The probability that Bob
gives outcome $j$ is therefore simply obtained by normalizing
Eq.~(\ref{13}), giving $(|b_j| + \vec a \cdot \vec b_j )/ 2$,
in accordance with the predictions of quantum mechanics.
Finally, since the probability of termination is 1/2,
the average number of iterations needed before acceptance by Bob is 2. 
The average number of bits transmitted is therefore twice the number
of bits exchanged during one iteration, the latter consisting of 2.19~bits
sent by Alice to Bob plus 1~bit sent by Bob to Alice. 
Thus, we obtain a total of $2(2.19 + 1) = 6.38$ bits.
\par

In summary,
we have exhibited a classical scenario based on a LHV model 
of quantum mechanics, which simulates the teleportation 
of an arbitrary (but known) qubit by use of classical
communication. It requires only 2.19~bits of communication
if Bob must be able to simulate any von Neumann measurement (or 
6.38~bits if Bob must simulate any POVM measurement).
This reflects that local hidden-variable models are in some sense 
surprisingly ``close'' to quantum mechanics: a very little amount of
classical communication is necessary to fill the gap between them, at least 
for low-dimensional systems. As the dimensionality increases, however,
the amount of communication increases exponentially (see \cite{BCT99}).

We should emphasize that, even though we have found an explicit method
that requires 2.19~bits, it is probably not optimal. Thus, the
minimum amount of classical communication needed for the classical
teleportation of a qubit may even be less than the amount necessary for 
quantum teleportation. Still, our protocol is surprisingly efficient
given that no prior entanglement is available. 
Indeed, if we restrict to 2 the average number of communicated bits 
per simulation with our scheme, the teleportation
fidelity that is attained
equals $(2+0.19 \times 0.5)/2.19 = 0.957$ (the fidelity 0.5
is achievable with LHVs only). This fidelity should be
compared to the 87\% fidelity achievable without Alice and Bob sharing 
LHVs but still exchanging 2 classical bits~\cite{gisin}.
\par

Another issue raised by our study concerns the interpretation of
the recent experimental realizations of teleportation\cite{teleportation}. 
Notwithstanding the fact that these
experiments are remarkable verifications of quantum mechanics, 
our result shows that one should be cautious when
asserting that such realizations use
the entangled pair in a way that could {\em not} be simulated by 
classical means. If the state that must be 
teleported is {\em unknown} to Alice, then there are bounds on the 
best quality of the transmission that can be obtained if only classical
communication is allowed and no entanglement is used. In contrast, 
we have shown that if the state is {\em known} to Alice, the transmission of 
the state can be simulated classically using only a few bits of
communication (in fact not a significantly larger amount than that
required for teleportation) if the two parties share local hidden variables. 
This remark will hopefully shed some new light on the interpretation of
these experiments.
\par

We are grateful to R. Cleve and B. Gisin for helpful discussions and
to R. Cleve for comments on the manuscript.
We thank the European Science Foundation for financial support. 
N.G. acknowledges funding by the Swiss National Science Foundation.
S.M. is {\it Chercheur qualifi\'e} of the Belgian National Fund for
Scientific Research.

\end{multicols}

\begin{references}
\bibitem{fn1} Such an infinite list of real numbers 
is actually no more information
than a single real number, since all the enumerable lists of real numbers
can be put in one-to-one correspondence with the real numbers themselves.
\end{references}
\end{document}